\documentclass{article}
\usepackage{paralist}
\usepackage{graphicx}
\usepackage{bm}
\usepackage{amsmath,amssymb} 
\usepackage{hyperref}
\usepackage{xcolor}
\usepackage{booktabs}
\usepackage{comment}
\usepackage{caption}
\usepackage{float}
\usepackage{threeparttable}
\usepackage{ragged2e}
\usepackage[sorting=none]{biblatex}
\usepackage[a4paper, total={6in, 8in}]{geometry}
\usepackage{setspace} 
\DeclareMathOperator{\sgn}{sgn}

\title{\bf Piezomagnetic switching of anomalous Hall effect in an antiferromagnet at room temperature}

\author{M. Ikhlas$^{1,2,3,\ast}$, S. Dasgupta$^{4,2,\ast}$, F. Theuss$^{5}$, T. Higo$^{1,2,3}$, Shunichiro Kittaka$^{6}$,\\ 
B. J. Ramshaw$^{5}$, O. Tchernyshyov$^{7}$, C. W. Hicks$^{8,9}$ \& S. Nakatsuji$^{1,2,3,7,10,\dagger}$}

\begin{document}

\maketitle

\begin{enumerate}[\expandafter\textsuperscript 1]
 \item Department of Physics, The University of Tokyo, Tokyo  113-0033, Japan
 \item Institute for Solid State Physics, The University of Tokyo, Kashiwa, Chiba 277-8581, Japan
 \item CREST, Japan
Science and Technology Agency (JST), Saitama 332-0012, Japan
 \item Department of Physics and Astronomy and Stewart Blusson Quantum Matter Institute, University of British Columbia, Vancouver,  British Columbia V6T 1Z1, Canada
 \item Laboratory of Atomic and Solid State Physics, Cornell University, Ithaca, NY 14853, USA 
  \item Department of Physics, Chuo University, Tokyo 112-8551, Japan 
 \item Institute for Quantum Matter and Department of Physics and Astronomy, Johns Hopkins University, Baltimore, MD 21218, USA
 \item School of Physics and Astronomy, University of Birmingham, Birmingham B15 2TT, UK
 \item Max Planck Institute for Chemical Physics of Solids, 01187 Dresden, Germany
 \item Trans-scale Quantum Science Institute, University of Tokyo, Tokyo  113-0033, Japan
 \item[$^\ast$] These authors contributed equally
 \item[$^\dagger$] Corresponding author
\end{enumerate}

\justifying
\textbf{Piezomagnetism couples strain linearly to magnetic order producing magnetization. Thus, unlike magnetostriction, it enables bidirectional control of a net magnetic moment. If this effect becomes large at room temperature, it may be technologically relevant, similar to its electric analogue, piezoelectricity. To date, however, the studies of the piezomagnetic effect have been primarily restricted to antiferromagnetic (AF) insulators at cryogenic temperatures. Here we report the discovery of a large piezomagnetism in a metal at room temperature. Strikingly, by using the AF Weyl semimetal Mn$_3$Sn, known for its nearly magnetization-free anomalous Hall effect (AHE), we find that an application of small uniaxial strain of the order of 0.1 \% can control both the sign and size of the AHE. Our experiment and theory show that the piezomagnetism can control the AHE distinctly from the magnetization, which will be useful for spintronics applications.}\\
\\
Application of strain causes dramatic changes in the physical properties of magnets; it can modulate the magnetic anisotropy, tune magnetic transition temperatures, and control multiferroic properties.\cite{ramesh2010multiferroics, dong2015multiferroic, song2017recent} In antiferromagnets with certain types of magnetic order, an application of strain breaks the symmetry between the antiferromagnetic (AF) sublattices and induces a net magnetic moment proportional to the applied strain.\cite{Dzialoshinskii:1958, borovik1994piezomagnetism,lukashev2008theory,zemen2017piezomagnetism, jaime2017piezomagnetism}  This phenomenon known as piezomagnetism, could be technologically relevant, similar to its electric analogue, piezoelectricity. It allows control over the magnitude and direction of the induced magnetization---in contrast to magnetostriction, which does not discriminate between opposite magnetization directions.\cite{lee1955magnetostriction} Experimental studies on the piezomagnetic effect have so far been confined mostly to insulators at cryogenic temperatures, and only very recently have metallic systems been investigated.\cite{borovik1994piezomagnetism,jaime2017piezomagnetism,boldrin2018giant,you2020room}  To date, the piezomagnetic effect has been utilized to manipulate AF domains not only in static conditions,\cite{radhakrishna1978neutron,baruchel1988piezomagnetism} but also dynamically at ultrafast time scales,\cite{disa2020polarizing} which can  potentially be useful for AF spintronics.\cite{Jungwirth2016,Baltz2018,Smejkal2018, liu2019antiferromagnetic} Here we show an empirical observation of a large piezomagnetic effect in the  Weyl antiferromagnet Mn$_{3}$Sn at room temperature, a material that has gained much attention for its promising use in spintronics devices.\cite{Kimata2019,Tsai2020,takeuchi2021chiral} Significantly, we demonstrate that the piezomagnetic effect can control not only the size but also the sign of the anomalous Hall effect (AHE) in Mn$_{3}$Sn.

Mn$_{3}$Sn has a hexagonal D0$_{19}$ structure (space group P6$_{3}$/$mmc$) comprising $[0001]$-axis stacked kagome lattice layers of Mn atoms with a Sn site at the center of a Star of David. We use a right-handed coordinate system with axes  $x$, $y$, and $z$ to represent the $[2\bar{1}\bar{1}0]$, $[01\bar{1}0]$, and $[0001]$ directions of the crystal structure.  Below its ordering temperature of $T_{\rm N} \sim 430$ K, the Mn moments form a 120$^\circ$ ``antichiral'' magnetic order shown in Fig.~1a.\cite{Nagamiya1982} Magnetic moments $\mathbf m_i$ on sublattices $i = 1,2,3$ of  magnitude  $m_i \approx 3\mu_B$ are oriented in the $xy$ plane and cancel each other nearly perfectly, adding up to a spontaneous net moment $\mathbf{M}_{\textnormal{S}} = \sum_{i=1}^3\mathbf m_i$ with a magnitude of order $10^{-3} m_i$. Neglecting this small net \emph{dipolar} magnetic moment, the magnetic unit cell can be viewed as a magnetic \emph{octupole}.\cite{Suzuki:2017} The \emph{octupolar} order parameter, under local $D_{3h}$ point group operations, reduces to a vector order parameter in the $xy$ plane \cite{dasgupta2020,Chen:2020}:
\begin{equation}
\mathbf K = K(\cos \Phi_K, \sin \Phi_K).
\label{eq:K-def}
\end{equation}
In the ground state, $\mathbf{K}$ is parallel to the direction of the Mn moment that points along the local easy axis $\mathbf{\hat{n}}$ (Fig.~\ref{fig:magnetic-order}a). For example, in the ground-state configuration illustrated in Fig.~\ref{fig:magnetic-order}a,  $\mathbf{K}$ points in the $+x$ direction.  We illustrate local magnetic states of Mn$_3$Sn as points on a clock in Fig.~\ref{fig:magnetic-order}b. When the moments are collectively rotated by angle $+\phi$, $\mathbf{K}$ rotates by $-\phi$. The odd-numbered states (I, III, ...) are ground states, while the even-numbered states are at energy maxima. In the absence of perturbations, the magnetization $\mathbf{M} \parallel \mathbf{K}$ in all the states. However, we shall see that in the presence of both strain and external magnetic field, their directions strongly diverge.

The $\mathcal{T}$-symmetry breaking octupolar order generates Weyl nodes near $E_{\rm F}$,\cite{kuroda2017evidence,yang2017topological,Chen2020} leading to the magnetization-free, large transverse responses---AHE, anomalous Nernst effect, and magneto-optical effects.\cite{nakatsuji:2015,ikhlas2017large,li2017anomalous, higo2018large, Matsuda2020} On the other hand, the small residual dipolar moment $\mathbf{M}_{\textnormal{S}}$ provides a negligible perturbation to the electronic structure.\cite{Suzuki:2017,yang2017topological,higo2018large}   This broken $\mathcal{T}$-symmetry also allows finite piezomagnetic coupling. From its magnetic point group symmetry, the piezomagnetic tensor $\Lambda$ of the antichiral phase expresses a proportionality between the components of in-plane magnetization and strain (Supplementary).  Microscopically, under uniaxial stress, the bond lengths, and hence the nearest-neighbor exchange interactions, become anisotropic. The rotation of the sublattice moments in response generates a net dipole moment. Since the initial spontaneous magnetization of a domain is tiny, a  small strain can invert its direction (Fig.~\ref{fig:magnetic-order}c). Consequently, under a constant magnetic field in the $xy$-plane, the piezomagnetic effect can mediate 180$^\circ$ domain reversal. For an unstrained sample, a positive and negative magnetic field along $\hat{\bf{x}}$ select state III and IX, respectively. In the presence of piezomagnetism, this situation is reversed under the appropriate direction and magnitude of strain, as illustrated in the free energy diagram in Fig.~\ref{fig:magnetic-order}d. This causes a sign change of the AHE under the combination of field and strain.

We first demonstrate the large piezomagnetism in Mn$_3$Sn. We measure the uniaxial stress dependence of the magnetization of two single-crystals grown by the Bridgman method, labelled M1 and M2, using a Cu-Be piston-cylinder cell (Methods). The samples have compositions of Mn$_{3+x}$Sn$_{1-x}$, ($x = 0.01-0.02$), and exhibit a first-order transition to an incommensurate state on cooling below $T_{\rm H} \sim 270-280$ K at ambient pressure, which is accompanied by a sharp decrease in the magnetization and the full suppression of the AHE.\cite{kren1975study, li2017anomalous, song2020complicated} Uniaxial stress $\sigma$ is converted to strain $\varepsilon$ using elastic constants determined by resonant ultrasound spectroscopy (Methods). 

Fig.~\ref{fig:Magnetization-stress}b shows the longitudinal magnetization $M$ vs. magnetic field $\mu_0 H$ of samples M1 and M2 under compressive stresses $\sigma_{xx}$ and $\sigma_{yy}$, respectively. At zero stress, M1 exhibits a spontaneous magnetization of $M_{\textnormal{S}}({\sigma=0})$ $\sim 3.6$ m$\mu_{\rm B}$/f.u., which can be flipped by a small coercive field of $\mu_{0} H_{c} \sim 0.03$ T, indicating a weak in-plane anisotropy. Furthermore, $M$ increases linearly with the field, with anisotropic slopes of $\chi = (\partial M/\partial H_x) = 14.3$ m$\mu_{\rm B}$/f.u. T$^{-1}$ along $x$ and $\chi = (\partial M/\partial H_y) = 13.5$  m$\mu_{\rm B}$/f.u T$^{-1}$ along $y$.  
Significantly, up to a threefold increase in $M_{\textnormal{S}}$ to 10.7 m$\mu_{\rm B}$/f.u. is obtained by  elastic compression of $\sigma_{xx} = -270$ MPa, before the sample fractures for $-\sigma_{xx} >  300$ MPa. Except for the increase in $M_{\textnormal{S}}$, no change is seen in the shape of the hysteresis curve and the susceptibility $\chi$. This confirms that below the yield stress the sample quality did not degrade, and that $M$ represents the magnetization of a single magnetic domain. The temperature scan measurements (Fig.~\ref{fig:Magnetization-stress}c) show that the stress-induced enhancement of $M$ only occurs in the antichiral phase ($T>T_{\rm H}$). 

As crucial evidence for piezomagnetism, the $\sigma$ dependence of $M_{\textnormal{S}}$ is found to be linear for both samples and nearly isotropic in the $xy$-plane for Sample M2 (Fig.~\ref{fig:Magnetization-stress}d). We fit the data with:
\begin{equation}
    M_{\textnormal{S}} = M_{\textnormal{S}}(\sigma =0) + \Lambda_{11} \sigma_{xx},  
\end{equation}
 where $\Lambda_{11}$ is the relevant piezomagnetic coefficient for the experimental configuration of Sample M1 in Fig.~\ref{fig:Magnetization-stress}b.  We obtain $\lvert \Lambda_{11} \rvert$ of  $\sim0.027$ m$\mu_{\rm B}$/f.u. MPa$^{-1}$ ($0.078$ Gauss/MPa) at room $T$, which is larger than the reported values for other transition metal compounds (see Table.~\ref{tab:piezomagneticon}) such as CoF$_{2}$,  but smaller than the largest value reported to date, $\Lambda_{14} = 0.226$ Gauss/MPa at $T=2.5$ K, for UO$_2$.\cite{jaime2017piezomagnetism} The susceptibility $\chi$, importantly, is stress independent. Moreover, in the tensile region ($\sigma > 0$), which our apparatus cannot access, the stress-induced magnetization is expected to compensate the zero-stress magnetization at a critical stress of $\sigma_{c} \approx 120$ MPa ($\varepsilon_{c}\approx 0.1\%$) (Fig.~\ref{fig:Magnetization-stress}d). As shown below, the energy scale of $ \sigma_{c} $ is directly connected to the strength of the local anisotropy $\delta$, and determines the sign-switching point of the AHE.

Next, we show that the large piezomagnetism in Mn$_3$Sn leads to a striking effect in electrical transport---strain control of the AHE sign. Generally, the dominant intrinsic contribution to the AHE in magnetic conductors is associated with a fictitious field experienced by conduction electrons in momentum space.\cite{Nagaosa:2010,Xiao2010} This field can be quantified by the Hall vector $\mathbf G$ representing the net Berry curvature of the occupied electronic states in the Brillouin zone.\cite{Nagaosa:2010,Xiao2010} The Hall vector directly determines the intrinsic part of the anomalous Hall conductivity tensor: 
\begin{equation}
\sigma^\text{Hall}_{\alpha\beta} = \frac{e^2}{(2\pi)^2 \hbar} \epsilon_{\alpha\beta\gamma} G_\gamma.   
\label{eq:Hall-vector-conductivity}
\end{equation}
Pseudovectors $\mathbf G$ and $\mathbf M$ have the same symmetry properties, are linearly related, and often collinear.
In Mn$_{3}$Sn, however, it is $\mathbf K$, not  $\mathbf M$,  that induces the intrinsic Hall conductivity in the antichiral phase  at $T_{\rm H} < T < T_{\rm N}$.\cite{Suzuki:2017,higo2018large} Thus, $\mathbf K$ and $\mathbf G$ should be identified as parallel to each other and confined to the $xy$-plane (Fig.~\ref{fig:magnetic-order}b),\cite{Suzuki:2017} and hereafter we use $\mathbf K$ to label both.
 
To study AHE under a continuous change of strain, we employ a high-precision piezoelectric-based strain device\cite{Hicks2014piezoelectric} and Mn$_{3}$Sn single crystals with similar high quality as the magnetization samples.  In the following, the strain values have been corrected for the effect of epoxy deformation, and the zero-strain points are verified by a separate set of measurements (Methods and Supplementary). We begin by describing the effect of  $x$-axis uniaxial strain, $\varepsilon_{xx}$, on the Hall resistivity $\rho_{zx}$ at $T=300$ K using Sample H1 (Fig.~\ref{fig:Hall-strain}a).

Fig.~\ref{fig:Hall-strain}d shows the field dependence of $\rho_{zx}$ at various strains.  At zero strain $\varepsilon_{xx} = 0\%$, the Hall resistivity exhibits a sharp hysteresis curve with a spontaneous Hall component of $\rho_{zx} (H=0) = -3.85$ $\mu\Omega$ cm and coercivity of $\sim 0.04$ T. Under the tension $ \parallel x$-axis of $\varepsilon_{xx}=0.194\%$, the overall shape of the sharp hysteresis curve is maintained. Namely, the spontaneous Hall component retains the  negative sign at positive fields but is slightly enhanced by 10\% from its unstrained value to $\rho_{zx}( H=0) = -4.2$ $\mu\Omega$ cm. For $\varepsilon_{xx} < 0 \%$,  significant changes to the field response of the Hall resistivity are observed. The remnant  Hall resistivity, $\rho_{zx}( H=0)$, gradually decreases under compression and is completely suppressed for $\varepsilon_{xx} < -0.2 \%$. Furthermore, for $\varepsilon_{xx} < -0.06 \%$, $\rho_{zx}$($H$) develops a component that is linear in $\mu_0 H$. At negative strain of $\varepsilon_{xx} = -0.21 \%$, $\rho_{zx}$ increases linearly with a slope of $\mu_{0}^{-1} (\partial \rho_{zx}/\partial H) \sim 1$ $\mu\Omega$ cm T$^{-1}$ until it reaches a positive saturation value of $\rho_{zx} = +4$ $\mu\Omega$ cm  at $\mu_{0} H = +3.8$ T. This AHE sign change under a strain of $\sim 0.1\%$ is particularly striking, because first-principles calculations on several magnetic materials predicted that a strain in the order of $1\%$ is generally required to produce significant changes in the AHE.\cite{samathrakis2020tailoring, tian2021manipulating}

Apart from the AHE sign change, under strongly negative strain, the shape of the hysteresis curve  acquires the qualitative form of a ferromagnet magnetized along its hard axis. This implies that a hard-axis (easy-axis) is induced perpendicular to negative (positive) strain direction, analogous to the inverse magnetostrictive effect in ferromagnets,\cite{lee1955magnetostriction} which is exactly what happens according to our theoretical analysis. Similar changes are observed in $\rho_{zy}$ when strain is applied along $y$-axis in Sample H2 (Fig. \ref{fig:Hall-strain}b and  e). In contrast, $z$-axis strains $\varepsilon_{zz}$ has negligible effect on $\rho_{zx}$ (Methods), which shows that $\mathbf{K}$ couples only to  $\varepsilon_{ij}(i,j=x,y)$. 

To examine the sign change of AHE in detail, we perform the Hall resistivity measurements with continuous change of strain, under a fixed magnetic field of $\mu_{0} H=6$ T perpendicular to the electric current and strain.  Fig.~\ref{fig:Hall-strain}f shows the strain dependence of the Hall resistivity for three representative samples normalized by their zero-strain values, $\rho_{\rm H}/\rho_{\rm H}\rvert_{\varepsilon=0}$, at $T=300$ K. Tensioning the sample enhances $\rho_{\rm H}/\rho_{\rm H}\rvert_{\varepsilon=0}$ slightly, from +1;  $\rho_{\rm H}/\rho_{\rm H}\rvert_{\varepsilon=0}$ to +1.18 at $\varepsilon =0.2\%$ in Sample H2.  The most noticeable features appear in the compression measurements. For all the samples, $\rho_{\rm H}/\rho_{\rm H}\rvert_{\varepsilon=0}$ decreases at negative strain and changes its sign at the switching strain $\epsilon_{switch}$ of  $\sim -0.075 \pm 0.02 \%$ ($\sigma_{switch}$ of $\sim 90 \pm 24$ MPa). This is approximately the same magnitude as the critical stress $\sigma_{c}$ inferred in the magnetization measurements, suggesting that $\sigma_{switch}$ and $\sigma_{c}$ are related. For $\epsilon < -0.2\%$, $\rho_{\rm H}/\rho_{\rm H}\rvert_{\varepsilon=0}$ asymptotically approaches a value around $-1$.  Additionally, $\rho_{\rm H}/\rho_{\rm H}\rvert_{\varepsilon=0}$ reversibly changes its sign under strain cycles, indicating that the samples remained elastic up to $-0.3 \%$ strain ($\sim 360$ MPa)---in contrast with the magnetization sample which fractured for $-\sigma > 300$ MPa. This is consistent with the previous studies which demonstrate that stress can be applied more homogeneously through epoxy in comparison to anvil-based cells.\cite{Hicks2014piezoelectric} Similar to the stress-induced enhancement of the magnetization, the strain-induced sign reversal of AHE only occurs for $T> T_{\rm H}$ (Fig.~\ref{fig:Hall-strain}g), which shows that both phenomena are intrinsic to the antichiral phase.

The piezomagnetism, and the reversal of AHE are captured in a Landau theory for the order parameter $\mathbf{K}$ (Supplementary). Deep in the ordered phase, the Hall vector has a well-defined magnitude $K$ and can be parametrized by its azimuthal angle $\Phi_K$ in Eq.~(\ref{eq:K-def}). Its coupling to an in-plane magnetic field, $\mathbf{H} = H(\cos{\Phi_H},  \sin{\Phi_H})$, and to shear strain, $(\varepsilon_{xx} - \varepsilon_{yy}, 2\varepsilon_{xy}) = \epsilon (\cos{2\Phi_\epsilon}, \sin{2\Phi_\epsilon})$, are restricted by the symmetries of rotation and time reversal. The energy terms bilinear in the local anisotropy strength $\delta (> 0)$, magnetic field $H$, and shear strain $\epsilon$ are:
\begin{equation}
U = 
- \delta H K \cos{(\Phi_K - \Phi_H)}
+ \delta \epsilon K^2 \cos{(2\Phi_K - 2\Phi_\epsilon)} 
+ \epsilon H K \cos{(\Phi_K + \Phi_H - 2 \Phi_\epsilon)}.
\label{eq:Landau-couplings}
\end{equation}
The first term here is the Zeeman coupling $- \mathbf H \cdot \mathbf M$ to the weak ferromagnetic moment $\mathbf M = \delta \mathbf K$ \cite{Liu:2017}, the second term represents uniaxial anisotropy induced by strain, and the last term describes piezomagnetism.\cite{Dzialoshinskii:1958} For brevity, material constants and $K$ in Eq.~(\ref{eq:Landau-couplings}) are set to unity by an appropriate choice of units for the energy density, field, and strain. 

Owing to the piezomagnetic term,  $M_{\textnormal{S}}$ changes linearly with in-plane uniaxial strain. For the configuration of Fig.~\ref{fig:Magnetization-stress}a, the magnetization along $\mathbf{H}$, per formula unit is (Supplementary):
\begin{equation}
    M = \frac{\hbar\gamma S}{2J}(\delta - \epsilon) + \frac{\hbar^{2}\gamma^{2}}{2J}\mu_{0} H,
    \label{eq:magnetization2}
\end{equation}
where the first (last) term is the spontaneous $M_{\textnormal{S}}$ (paramagnetic) component. For  tensile strain of $\epsilon = \delta$, the strain-induced magnetization compensates the zero-strain magnetization. Note that the susceptibility $\chi = \partial M/\partial H$ is independent of strain, as observed in experiment. Assuming a local moment of $S=3/2$  and the gyromagnetic ratio  of $\gamma = 2\mu_{B}/\hbar$,\cite{Cable_1993a} the fitting of Eq.~(\ref{eq:magnetization2}) to our results shown in Fig.~\ref{fig:Magnetization-stress} yields the parameters $d\epsilon/d\sigma \sim 0.155$ meV/GPa, $J \sim 8.6$ meV, $\delta \sim  0.02$ meV; the latter two parameters are consistent with neutron scattering experiments.\cite{Park2018} 

Let us now examine the strain dependence of $\mathbf{K}$. The free energy Eq. (\ref{eq:Landau-couplings}) simplifies to: 
\begin{equation}
U = 
- (\delta + \epsilon) H K_\gamma  
- 2\delta \epsilon K_\gamma^2,
\label{eq:Landau-couplings-Kx}
\end{equation}
for the configuration in Fig.~\ref{fig:Hall-strain}b (Fig.~\ref{fig:Hall-strain}a) with $K_\gamma = K_x (K_y)$ restricted to the interval between $\pm 1$. The location of the free-energy minimum depends on the relative strengths of $\delta$, $H$, and $\epsilon$. Notably, our theory not only reproduces the observed sign change in AHE but identifies three distinct regimes, in which the behavior of $K_x$ as a function of $H$ resembles the magnetization curves in a ferromagnetic, paramagnetic, and diamagnetic material. We term these regimes \emph{ferrohallic}, \emph{parahallic}, and \emph{diahallic}.

For positive shear strain, $\epsilon>0$, the quadratic term in Eq.~(\ref{eq:Landau-couplings-Kx}) is negative and the minimum of the free energy lies at the ends of the interval $-1 \leq K_x \leq +1$. An arbitrarily weak magnetic field breaks the symmetry and selects $K_x = \sgn(H)$, as shown in Fig.~\ref{fig:Kx-curves}a. For negative shear strain, $\epsilon < 0$, the sign of the quadratic term in Eq.~(\ref{eq:Landau-couplings-Kx}) turns positive and the minimum of the free energy shifts inside the allowed interval $-1 \leq K_x \leq +1$ below the saturation field $H_s = |\frac{4\delta\epsilon}{\delta+\epsilon}|$. $K_x$ increases linearly with $H$ before saturating at $K_x = \pm 1$ for $|H|\geq H_s$. The slope $dK_x/dH = -(\delta +\epsilon)/4\delta\epsilon$ is positive (parahallic) for moderately negative strain, $-\delta < \epsilon < 0$, Fig.~\ref{fig:Kx-curves}b), and negative (diahallic) for strongly negative strain, $\epsilon < -\delta$, (Fig.~\ref{fig:Kx-curves}c) before saturating at $K_x = \sgn(H)$ and $K_x = -\sgn(H)$, respectively.

It is instructive to examine the shear strain dependence of $\mathbf K$ in a fixed magnetic field. Three distinct regimes are also seen (Fig.~\ref{fig:Kx-curves}d). For $\epsilon \geq 0$, we find $K_x = \sgn{(\rm H)}$. As the strain goes down below an upper critical strength $\epsilon_+ = -\frac{H\delta}{H + 4\delta} < 0$, $K_x$ begins to decrease and crosses 0 at $\epsilon_c = - \delta$. If the field is weak, $H<4\delta$, $K_x$ descends toward an asymptotic value, $-H/4\delta$. For strong fields, $H > 4\delta$, $K_x$ reaches $-1$ at the lower critical strain $\epsilon_- = -\frac{H\delta}{H - 4\delta}$ and saturates at $K_x = - \sgn(H)$. Intuitively, the critical strain $\epsilon_c = -\delta$ is the point at which the direction of $M_{\textnormal{S}}$ of a domain inverts. Importantly, our theory enables the quantitative estimate of parameters using the results of the Hall effect measurements (Fig.~\ref{fig:Kx-curves}c \& d). They are found consistent with the corresponding values estimated from the magnetization measurements (Supplementary).

The theoretical analysis yields another important implication. Without strain, the magnetization is parallel to the Hall vector, $\mathbf M = \delta \mathbf K$, but this simple relation breaks down in the presence of applied strain due to the large piezomagnetic effect. While $\mathbf K$ can be rotated $180^\circ$ away from the direction of the field by sufficiently strong strain, the Zeeman coupling ($-\mathbf H \cdot \mathbf M$) ensures that the magnetization always has a component aligned with the magnetic field, even in the presence of anisotropy (Methods). Indeed, as can be seen in Fig.~4{a}-{c}, the sign of the longitudinal magnetization is fixed, while $K_{x}$ changes sign. Thus, by applying both strain and external magnetic field, we can separately control the underlying order parameter $\mathbf K$ from the auxiliary magnetic dipole moment $\mathbf{M}$. Illustrations of this is shown in Supplementary Fig.~7 (Supplementary). 

Finally, our strain control of AHE  engenders additional means to control antiferromagnets, complementary to magnetic field and electrical current.\cite{Tsai2020,nakatsuji:2015, Baltz2018} In thin films, a large strain can be applied by lattice mismatch from substrate or by using a piezoelectric material.\cite{ramesh2010multiferroics, dong2015multiferroic,song2017recent, liu2019antiferromagnetic}  Given the recent report on the gigantic THz optical enhancement\cite{disa2020polarizing} as well as the perspective of AF spintronics,\cite{Jungwirth2016,Baltz2018,Smejkal2018} the piezomagnetic effect  may become useful in facilitating the ultrafast operation of antiferromagnets.


\bigskip
\noindent
\textbf{Acknowledgments:} The work at the Institute for Quantum Matter, an Energy Frontier Research Center was funded by DOE, Office of Science, Basic Energy Sciences under Award \# DE-SC0019331. This work was partially supported by JST-Mirai Program (JPMJMI20A1), JST-CREST (JPMJCR18T3), JST-PRESTO (JPMJPR20L7), Japan Science and Technology Agency, Grants-in-Aids for Scientific Research on Innovative Areas (15H05882, 15H05883, 15K21732) from the Ministry of Education, Culture, Sports, Science, and Technology of Japan, and Grants-in-Aid for Scientific Research (19H00650). S.N. acknowledges support from the CIFAR as a Fellow of the CIFAR Quantum Materials Research Program. The use of the facilities of the Materials Design and Characterization Laboratory at the Institute for Solid State Physics, The University of Tokyo, as well as the Cryogenic Research Center, The University of Tokyo, is gratefully acknowledged.  M.I. is supported by a JSPS Research Fellowship for Young Scientists (DC1). S.D. is supported by funding from the Max Planck-UBC-UTokyo Center for Quantum Materials, the Canada First Research Excellence Fund, Quantum Materials and Future Technologies Program, and the Japan Society for the Promotion of Science KAKENHI Grant No. JP19H01808. CWH acknowledges support from the Deutsche Forschungsgemeinschaft through SFB 1143 (Project ID 247310070) and the Max Planck Society. The identification of any commercial product or trade name does not imply endorsement or recommendation by the National Institute of Standards and Technology. \\ \\
 \noindent
\textbf{Author Contributions :} S.N. and O.T. conceived the project; S.N., B.J.R and C.W.H planned and supervised the experiments; M.I. synthesised and prepared the samples; M.I. and T.H. performed the transport measurements under uniaxial strain and magnetization measurements under uniaxial stress; M.I. performed the finite element simulations; F.T. and B.J.R. conducted the resonant ultrasound spectroscopy measurements; S. K. developed the piston-cylinder type pressure cell; C.W.H. developed the uniaxial strain cell; S.D. and O.T. developed the Landau theory and S.D. performed the numerical calculations; M.I., S.D., S.N., and O.T. wrote the manuscript with the comments from F.T., C.W.H.. All authors read and commented on the manuscript. \\ \\
\noindent
\textbf{Competing Interests :} The authors declare that they have no competing financial interests.\\ \\
\noindent
\textbf{Correspondence :} Correspondence and requests for materials should be addressed to Satoru Nakatsuji (email: satoru@phys.s.u-tokyo.ac.jp).


\newpage


 \begin{table}[h]
	 \caption{{\bf{Experimental piezomagnetic (PM) coefficients of selected bulk materials}}.}
	 \label{tab:piezomagneticon}
	 \centering
	 \begin{tabular}{l l l l }
		 \toprule
		 Compound & $T$ (K)  &   PM coeff. (Gauss/MPa) & Reference\\
		 \midrule
		 Mn$_3$Sn & 300   & $\Lambda_{11}=0.055$ & This work\\
		 \addlinespace
		 CoF$_{2}$ & 20  &  $\Lambda_{14}=0.021$; $\Lambda_{36}=0.008$ & \cite{borovik1960piezomagnetism}\\
		 \addlinespace
		 MnF$_{2}$ & 60   & $\Lambda_{36}=0.00087$ & \cite{baruchel1988piezomagnetism}\\
		 \addlinespace
		 $\alpha$-Fe$_2$O$_3$ & $\approx 80$ & $\Lambda_{11}=0.024$ & \cite{andratski1967zh, voskanyan1968magnetostriction}\\
		 \addlinespace
		 DyFeO$_3$ & 6   & $\Lambda_{36}=0.075$ & \cite{zvezdin1985linear} \\
		 \addlinespace
		 YFeO$_3$ & 6   & $\Lambda_{15}=0.010$ & \cite{kadomtseva1981direct} \\
		 \addlinespace
		 UO$_2$ & 2.5   & $\Lambda_{14}=0.226$ & \cite{jaime2017piezomagnetism} \\
		 \bottomrule\\
	 \end{tabular}
 \end{table}
\newpage
 


\begin{centering}
\includegraphics[width=1\columnwidth]{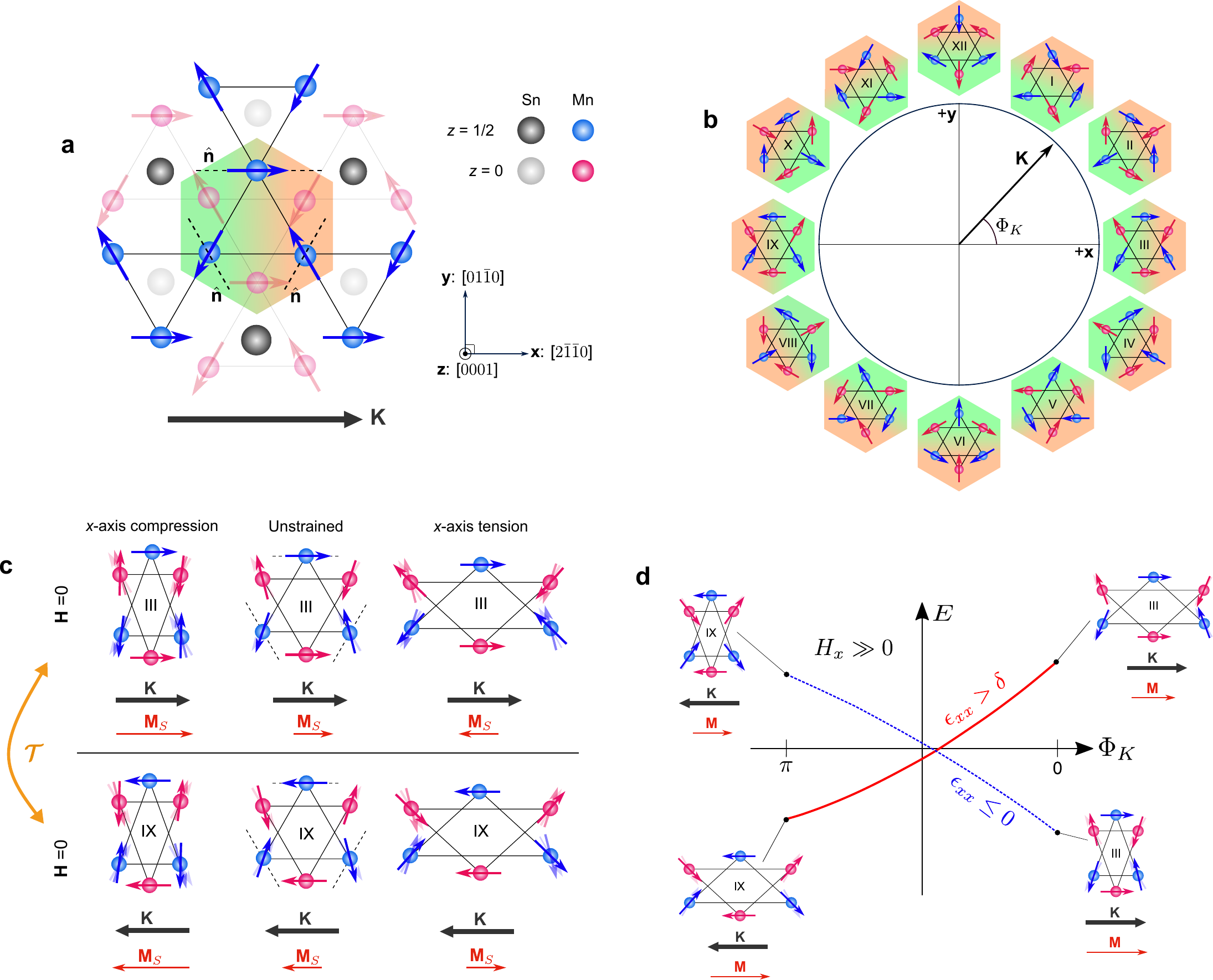}
\captionsetup{parbox=none}
\captionof{figure}{{\bf{120$^{\circ}$ antichiral magnetic structure of Mn$_{3}$Sn and piezomagnetic control of the magnetization ${\mathbf{M}}$} and the divergence of its direction from the order parameter ${\mathbf{K}}$}. ({\bf{a}}) Magnetic order in Mn$_3$Sn. The colored hexagon represents a single unit of magnetic octupole, consisting of two adjacent triangles from different kagome planes. Black dashed lines are local easy axes $\mathbf{\hat{n}}$ with an associated weak energy scale $\delta$. The octupole order parameter is represented by the vector ${\mathbf{K}}=K(\cos\Phi_{K},\sin\Phi_{K} )$ in the $xy$ plane.  ({\bf{b}}) Local magnetic states of Mn$_{3}$Sn, represented as points on a clock; the odd-numbered states are the ground states.  ({\bf{c}}) Illustration of piezomagnetic effect in Mn$_{3}$Sn. In the unstrained state, there is a small spontaneous magnetization $\mathbf{M}_{\textnormal{S}}(\parallel \mathbf{K})$, which arises from the tilting of the sublattice moments towards their respective local easy axes. Applied uniaxial stress makes the exchange interaction anisotropic. For example, in state III under $x$-axis compression, the horizontal bonds become shorter, the exchange interaction across these bonds increases, and the spins connected by horizontal bonds rotate to become more nearly anti-parallel. That rotation increases the net magnetic moment along $+x$. States related by time reversal operation $\mathcal{T}$ ($\pm{\mathbf{K}}$) have opposite strain-induced magnetization.  ({\bf{d}}) Illustration of the free energy $E$  as a function of the order parameter orientation $\Phi_{K}$ in the $xy$-plane, at a fixed magnetic field $\mu_0 H \gg 0$ parallel to +{\bf{x}}. For zero and negative $x$-axis strain ($\epsilon_{xx} \leq 0$, left panel), state III is the lowest energy state ($\Phi_{K}= 0$). In the presence of piezomagnetism, when $x$-axis tension is applied, with an associated energy greater scale than the local anisotropy ($\epsilon_{xx} > \delta$, right panel), state IX ($\Phi_{K}= \pi$) has the lowest energy. Here, $\mathbf{M}$ is the total magnetization, which includes both the spontaneous ($M_S$) and field-induced components. }
\label{fig:magnetic-order}
\end{centering}

\newpage

\begin{centering}
\includegraphics[width=1\columnwidth]{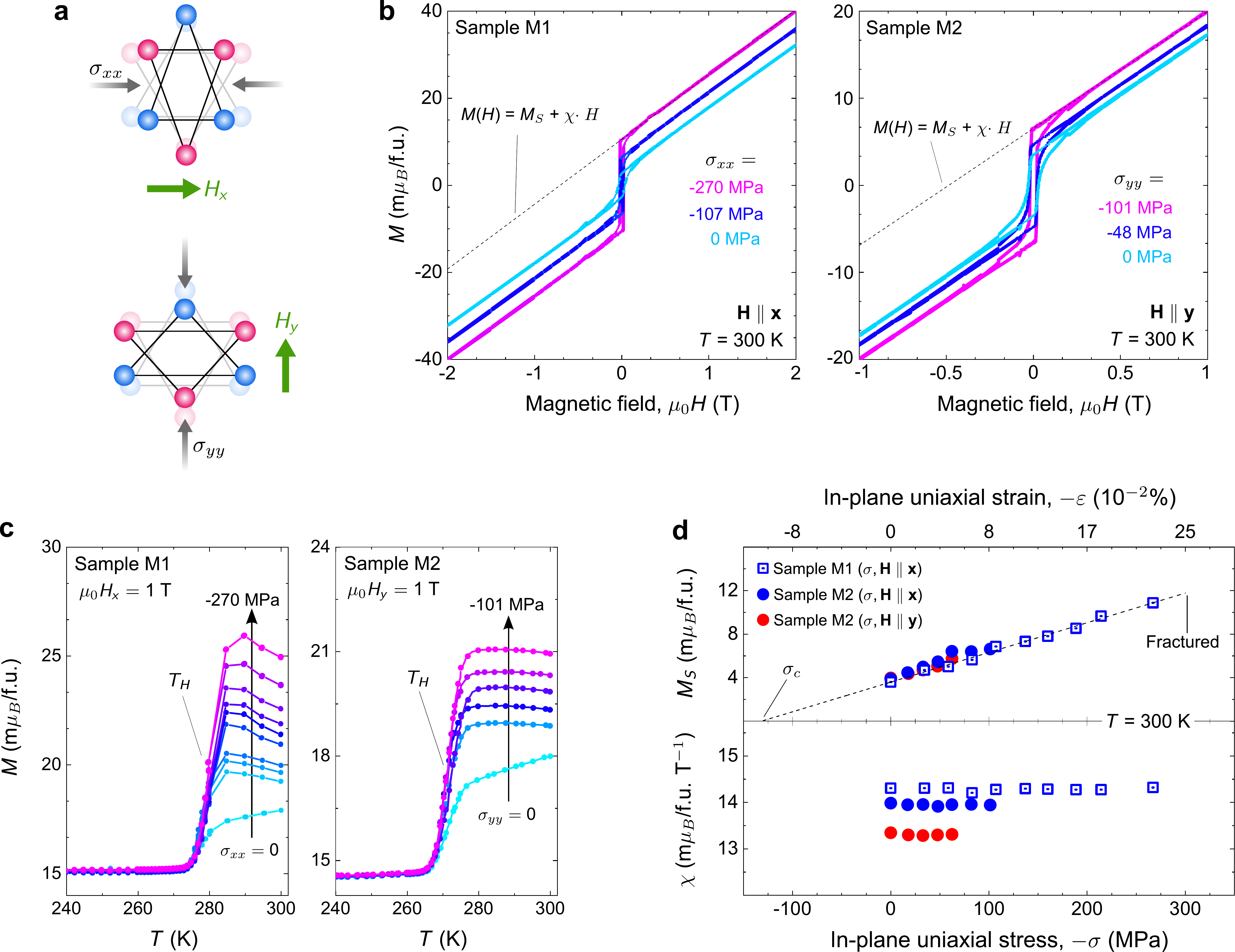}
\captionsetup{parbox=none}
\captionof{figure}{{\bf{Piezomagnetic effect in the topological antiferromagnet Mn$_{3}$Sn under in-plane uniaxial compression}}. ({\bf{a}}) Sample configurations for the magnetization measurements under uniaxial stress. ({\bf{b}}) Longitudinal magnetization $M$ vs. magnetic field $\mu_0 H$ at $T=300$ K under uniaxial stress $\sigma$ along the $x$-axis for Sample M1 (left panel) and along the $y$-axis for Sample M3 (right panel) . Dashed line represents a linear fit to the data in the region where the hysteresis loop closes ($\mu_0 H>0.5$ T). ({\bf{c}}) Temperature dependence of the magnetization under various uniaxial stress, measured on cooling from $T=300$ K. Left (Right) panel shows the data for Sample M1 (M3) under a magnetic field of $\mu_{0} H=1$ T and uniaxial stress $\sigma$ applied along the $x$-axis ($y$-axis).  $T_{\rm{H}}$ indicates the incommensurate transition temperature for the respective samples.  ({\bf{d}}) Uniaxial stress $\sigma$ dependence of the spontaneous magnetization $M_{\textnormal{S}}$ and magnetic susceptibility $\chi$ for Sample M1, M2, and M3 at $T=300$ K. Blue (Red) data points correspond to data taken under field and stress applied along the $x$($y$)-axis. The upper axis shows the corresponding uniaxial strain $\varepsilon$ inferred from  the experimental Young's modulus $E_{11}=121 $ GPa. Solid blue and red lines represent the linear fits to the $M_{\textnormal{S}}$ vs. $\sigma$ data of Sample M1 and M3, respectively. $\sigma_{c} \approx 120$ MPa is the critical uniaxial stress for which the stress-induced magnetization compensates the ambient pressure spontaneous magnetization $M_S (\sigma=0)$. Negative (positive) stress denotes compression (tension).}
\label{fig:Magnetization-stress}
\end{centering}

\newpage

\begin{centering}
\includegraphics[width=1\columnwidth]{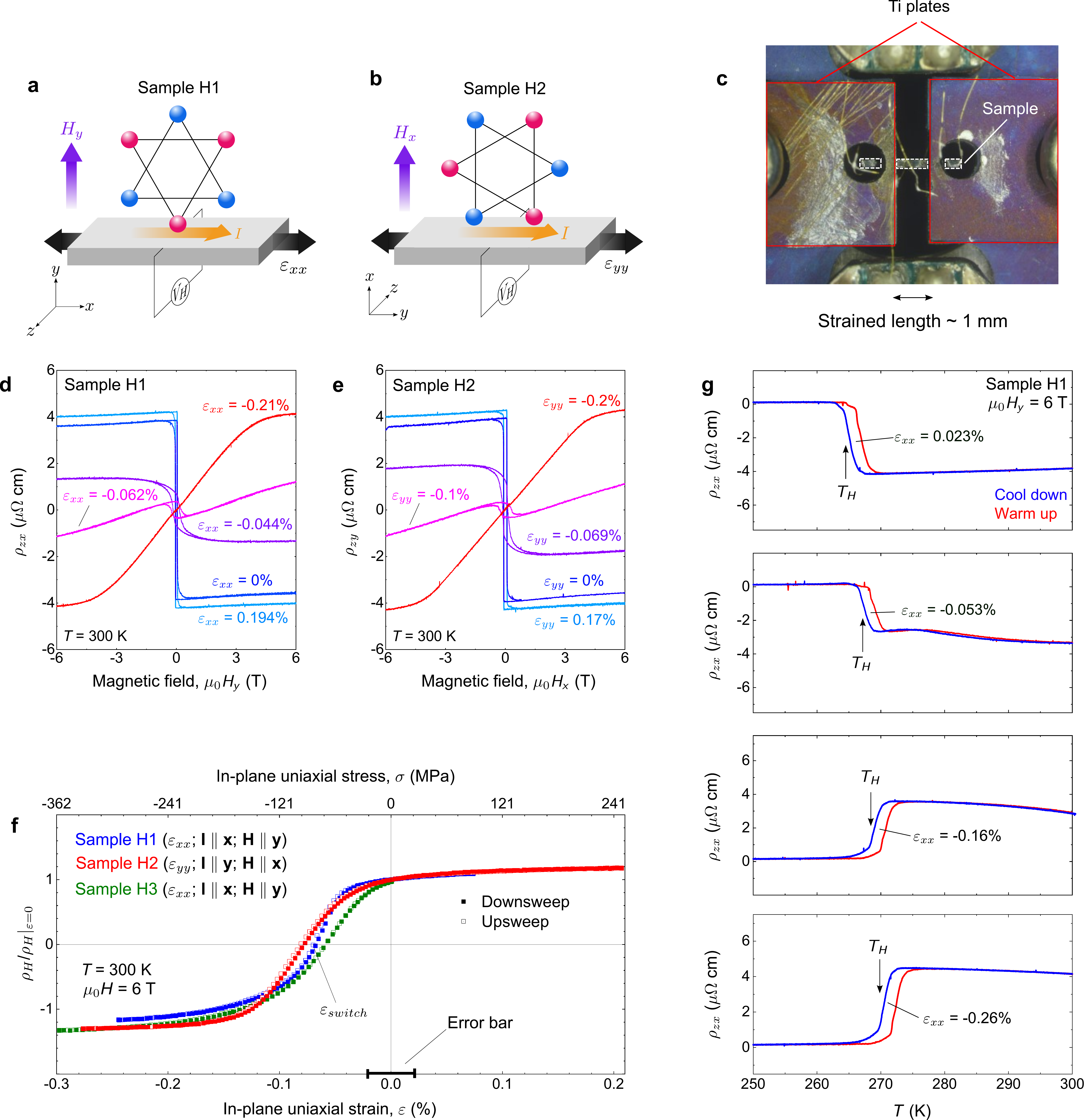}
\captionsetup{parbox=none}
\captionof{figure}{{\bf{Anomalous Hall effect of the Weyl antiferromagnet Mn$_{3}$Sn and its sign reversal under in-plane uniaxial strain}}. ({\bf{a, b}}) Hall measurement configuration for Sample H1 and Sample H2, respectively. ({\bf{c}}) The image of a sample mounted on the strain cell. ({\bf{d, e}}) Results of measurements in the configurations illustrated in panels {\bf{a}} and {\bf{b}}, respectively. ({\bf{f}}) In-plane uniaxial strain $\varepsilon$ dependence of the Hall resistivity normalized by its zero-strain value $\rho_{\rm H}/\rho_{\rm H}\rvert_{\varepsilon=0}$ for three representative samples, taken at $\mu_{0} H =6$ T and $T=300$ K. Sample H3 is measured using a special sample mounting method (methods). Closed squares represent data measured as $\varepsilon$ is swept from its highest tension value ($\varepsilon>0$)   to its highest compression value ($\varepsilon<0$), while open squares represent the data taken in the opposite scan direction. Upper axis shows the uniaxial stress value $\sigma$ corresponding to the applied strain, obtained using Young's modulus of $E_{11}$ = 121 GPa. Error bar of $\Delta \varepsilon = \pm 0.02\%$ is associated with the uncertainty in the strain transmission value (methods). $\varepsilon_{switch} \approx 0.075 \%$ is the point where AHE changes its sign. ({\bf{g}}) Temperature dependence of the Hall resistivity $\rho_{zx}$ of Sample H1 for various applied strains, taken at $\mu_{0} H=6$ T $\parallel y$-axis. Blue (red) curve corresponds to data taken when cooling down (warming up) between 250 K and 300 K. Arrows indicate the location of first-order incommensurate transition temperature $T_{\rm H}$. Negative (positive) strain denotes compression (tension).}
\label{fig:Hall-strain}
\end{centering}

\newpage

\begin{centering}
\includegraphics[width=1\columnwidth]{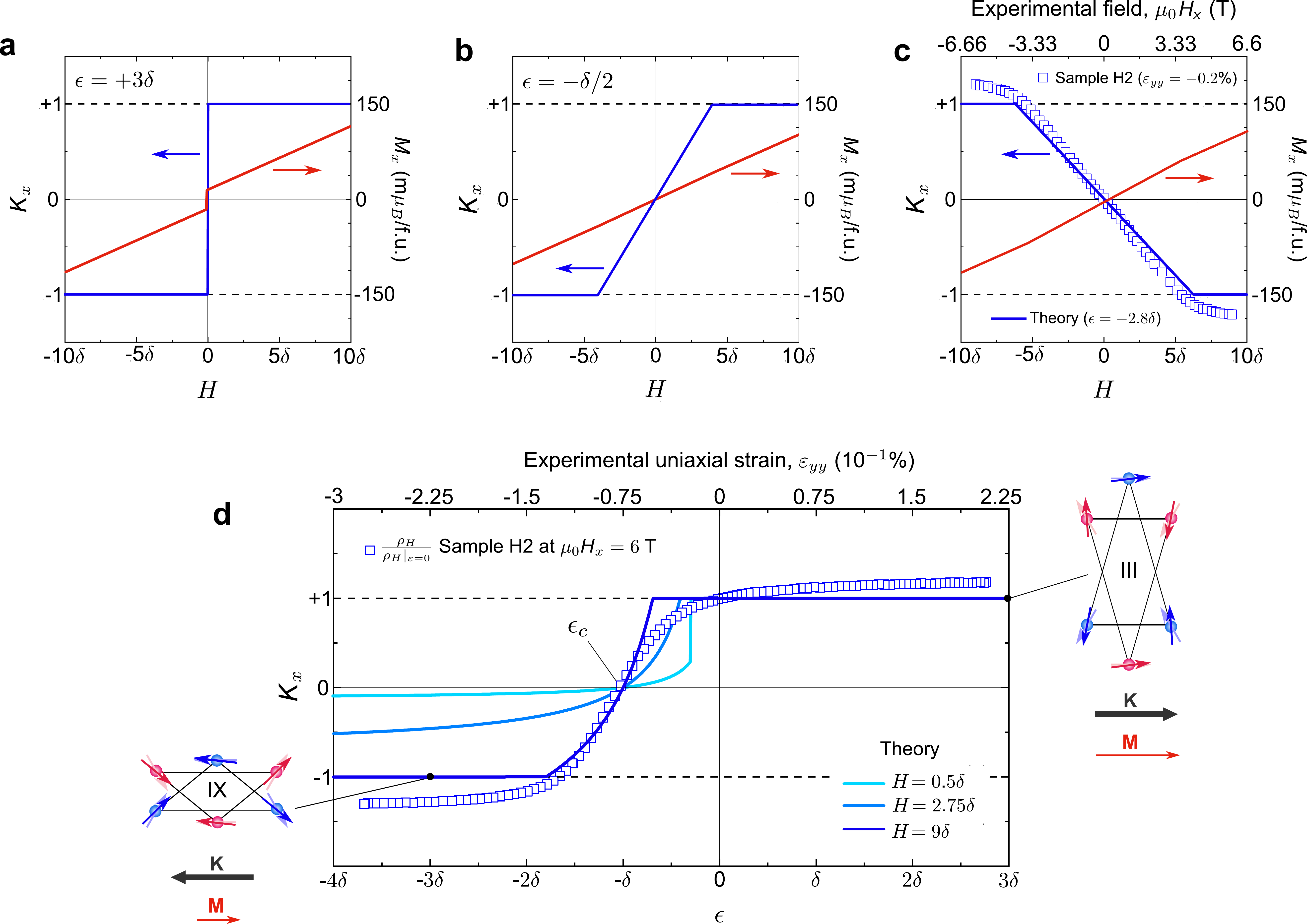}
\captionsetup{parbox=none}
\captionof{figure}{\textbf{Distinct strain controls of the Hall vector K under magnetic field in the ferrohalic, parahallic, and diahallic regimes.} ({\bf{a,b,d}}) Theoretical curves for the $x$-component of the Hall vector $K_x(H)$ (blue) and magnetization $M_{x} (H)$ (red) for fixed shear strain $\epsilon$ applied perpendicular to the field $H$ along {\bf{+x}} ($\Phi_{\epsilon} = \pi/2$; $\Phi_{H} = 0$) in the ({\bf{a}}) ferrohallic ($\epsilon \geq 0$), ({\bf{b}}) parahallic ($ -\delta < \epsilon < 0$), and ({\bf{c}}) diahallic regimes ($\epsilon < -\delta$). In {\bf{c}}, we show the experimental result (blue open squares) of the $T=300$ K Hall resistivity of Sample H2 at $\epsilon_{yy} = -0.2\%$, normalized by its zero strain value at $T=300$ K.  ({\bf{d}}) Theoretical curves for $K_x(\epsilon)$ at fixed magnetic fields $H$ applied perpendicular to shear strain $\epsilon$ ($\Phi_{\epsilon} = \pi/2$; $\Phi_{H} = 0$); Hall vector $K_{x}$ changes sign at a critical strain of $\epsilon_{c}= -\delta$. Additionally, we included the experimental data for Sample H2 at $T=300$ K and $\mu_{0}H_{x}=6$ T. All the curves are calculated with: $J=1$, $\delta=0.005$, $S=3/2$, $\gamma=2\mu_{B}/\hbar$, respectively. Schematic configurations of domains III and IX are shown with the parameter $3\delta$ and $-3\delta$, respectively; faded arrows represent the ideal 120$^{\circ}$ configuration, and the rotation of the sublattice moments from the ideal configuration are exaggerated by a factor of 25.}
\label{fig:Kx-curves}
\end{centering}

\newpage

\printbibliography[title={References}]

\newpage

\justifying
\noindent
{\Large \textbf{Methods}} \\\\
\noindent
\textbf{Sample synthesis and preparation} Single crystals of Mn$_{3}$Sn were synthesised from excess Sn using solution Bridgman method.\cite{caillat1996bridgman} First, we pre-melted Mn (Rare Metallics, 99.99\%) and Sn (Rare Metallics, 99.999\%) with a molar ratio of $2.97:1$ within a conical alumina crucible in an evacuated quartz ampoule at 1050$^{\circ}$C. Then, we transferred the precursor ingot to a two-zone vertical Bridgman furnace with a central temperature of $T=1100^{\circ}$C. Finally, the melt was passed through a temperature gradient of $\sim 6^{\circ}$C/mm with a growth speed of 0.25 mm/h. The resulting ingot contained two well-segregated sections: a lower section consisting of Mn$_{3}$Sn single crystals and a top section consisting of Sn-rich eutectic. The orientation of the Mn$_{3}$Sn crystals were checked using backscattering Laue X-ray diffraction (Photonic Science Ltd.) and their phase purity was confirmed using powder X-ray diffraction (Cu K$\alpha$, Rigaku Smartlab). We cut the crystals using spark erosion and polished them into bar-like shapes for transport, magnetization, and resonant ultrasound spectroscopy measurements. Chemical analysis performed using inductively coupled plasma optical emission spectroscopy (ICP-OES) indicates that the samples used in this study have compositions Mn$_{3+x}$Sn$_{1-x}$ in the range of $0.01<x<0.02$.\\\\
\noindent
\textbf{Resonant ultrasound spectroscopy} We determined the elastic moduli of Mn$_{3}$Sn using resonant ultrasound spectroscopy (RUS). A single crystal was polished to a rectangular prism with faces perpendicular to the $\mathbf{x}: [2\bar{1}\bar{1}0]$, $\mathbf{y}:[0\bar{1}\bar{1}0]$, and $\mathbf{z}:[0001]$ crystallographic directions. To perform RUS, the sample is held in weak-coupling contact between two piezoelectric transducers. The excitation frequency of one transducer is swept, while the response on the second transducer is detected via standard lock-in techniques.\cite{Ghosh2020} This gives us all of the mechanical resonances of the sample over a chosen frequency range. In this case, we measured 68 resonances between 950 kHZ and 4 MHz. From these resonance frequencies, we obtain the full elastic tensor by inverse solving the three dimensional elastic wave equation.\cite{Visscher1991,ramshaw2015avoided}  
 
There are four irreducible representations of strain in D$_{6\rm{h}}$---the point group of Mn$_3$Sn. This results in five independent elastic moduli C$_{ij}$: one for each strain and another coupling the two A$_{1\rm{g}}$ strains together. In Voigt notation, this reduces the stress-strain relationship, $ \mathbf{\sigma} = \mathbf{C} \mathbf{\varepsilon}$, to

 \begin{equation*}
 \begin{pmatrix}
 \sigma_{1} \\
 \sigma_{2} \\
 \sigma_{3} \\
 \sigma_{4} \\
 \sigma_{5}\\
 \sigma_{6}\\
 \end{pmatrix} = 
 \begin{pmatrix}
 C_{11} & C_{12} & C_{13} & 0 & 0 & 0\\
 C_{12} & C_{11} & C_{13} & 0 & 0 &0\\
 C_{13}  & C_{13}  & C_{33} & 0 & 0 & 0  \\
 0 & 0 & 0 & C_{44} & 0 & 0 \\
 0 & 0 & 0 & 0 & C_{44} & 0 \\
 0 & 0 & 0 & 0 & 0 & \left(C_{11}-C_{12}\right)/2\\
 \end{pmatrix}
 \begin{pmatrix}
 \varepsilon_{xx}  \\
 \varepsilon_{yy}  \\
 \varepsilon_{zz}  \\
 2\varepsilon_{yz} \\
 2\varepsilon_{xz} \\
 2\varepsilon_{xy} \\
 \end{pmatrix} .
 \end{equation*}
 
 \noindent Supplementary Table 1 gives the elastic moduli of Mn$_{3}$Sn at $T = 435$ K and $T=395$ K as measured by RUS. Additionally, we computed the Young's moduli and Poisson's ratios. They are most easily expressed in terms of the compliance tensor, given by the inverse of the elastic tensor $\mathbf{C}$,  $\mathbf{S} = \mathbf{C}^{-1} $. We can derive the Young's modulus along the [2$\bar{1}$$\bar{1}$0] and [0$\bar{1}$$\bar{1}$0] axis from $E_{11} =S_{11}^{-1}$, the Young's modulus along the [0001] axis from $E_{33} =S_{33}^{-1}$, the in-plane Poisson's ratio from $\nu_{xy} = -S_{12}/S_{11}$, and the out-of-plane Poisson's ratio from $\nu_{zx} = -S_{13}/S_{11}$. In the table, the errors represent uncertainty of 5 $\mu$m in the sample dimension, following Ghosh {\emph{et. al.}}.\cite{Ghosh2020}\\\\
 \noindent
 \textbf{Magnetization measurement under uniaxial compression}  We performed magnetization measurements under  uniaxial stress using piston-cylinder type pressure cell made of Cu-Be alloy with polybenzimidazole (PBI) cylindrical outer body.\cite{kittaka2010} The samples were shaped into cuboids with masses of $\sim 20$ mg, and their faces were directly put into contact with the piston. We estimated the uniaxial stress by dividing the force $F$ applied to the piston using hydraulic press at $T=300$ K by the cross sections $A$ of the samples, $\sigma = F/A $.   We carried out the measurements using a commercial system (Quantum Design MPMS) in the temperature range of $T= 240 - 300$ K and magnetic field up to $\pm$ 2 T parallel to the stress direction. To estimate the diamagnetic contribution of the cell, we subtracted the zero-stress magnetization $M (\sigma=0)$ data of the samples measured using the pressure cell with the $M (\sigma=0)$ data of the samples measured using ordinary translucent plastic straw. Because there is a differential thermal contraction between the cell body and the sample, the actual sample stress increases upon cooling from room temperature; we estimated this thermally-induced stress in the Supplementary. \\\\
 \noindent
 
\textbf{Transport measurements under uniaxial strain} We performed transport measurements under uniaxial strain using a custom piezoelectric-based strain cell (Supplementary Fig.~1a).  The samples' ends were epoxied between two titanium plates using degassed Loctite\textsuperscript{\textregistered} Stycast 2850FT; the strain is monitored by parallel plate capacitors and can be determined with higher accuracy than in a previous work.\cite{Hicks2014piezoelectric} We calculated the sample strain $\varepsilon$  using:
\begin{equation}
    \varepsilon = (\textrm{strain transmission}) \times \varepsilon_{0}A \frac{(1/C_{1} - 1/C_{0})}{L},
\end{equation}
\noindent where $\varepsilon_{0}$ is the vacuum permittivity, $A$ is the area of the parallel plates, $C_{0}$ is the sensor's baseline capacitance, $C_{1}$ is the sensor's capacitance after voltage is applied to the piezoelectric actuators, and $L$ is the sample's exposed length. Here, the strain transmission factor ($<1$) appears because strain is not perfectly transferred to the sample due to the deformation of the epoxy. We estimated the strain transmission and the uncertainty of the strain using finite element analysis (Supplementary). Negative and positive strains denote compression and tension, respectively. The value of $C_{0}$ varies between each measurement runs, and thus we define $C_{0}$ to be the sensor's initial capacitance at $T=300$ K before voltage is applied to the piezos. The capacitance of the sensor is temperature-dependent even at zero actuator voltage. Therefore, when we fix the capacitance setpoint at $T=300$ K by applying voltage feedback to the actuators, the sample strain will change as a function of $T$ due to the shift in the capacitance baseline, which is shown in Supplementary Fig. 1b. We took this baseline shift into account when estimating the strain in temperature scan measurements.


\end{document}